\documentclass[twocolumn]{aastex701}
\usepackage{physics,tabularx}
\usepackage{booktabs,makecell}

\newcommand{\rev}[1]{#1}

\begin{document}

\title{Water enrichment of forming sub-Neptune envelopes limited by oxygen exhaustion}

\author[orcid=0000-0001-8477-2523]{Tadahiro Kimura}
\affiliation{UTokyo Organization for Planetary Space Science (UTOPS), University of Tokyo, Hongo, Bunkyo-ku, Tokyo 113-0033, Japan}
\affiliation{Kapteyn Astronomical Institute, University of Groningen Landleven 12, 9747 AD, Groningen, Netherlands}
\email[show]{t.kimura624@gmail.com} 

\author[orcid=0000-0002-3286-7683]{Tim Lichtenberg} 
\affiliation{Kapteyn Astronomical Institute, University of Groningen Landleven 12, 9747 AD, Groningen, Netherlands}
\email{tim.lichtenberg@rug.nl}

\begin{abstract}
The interaction between a magma ocean and a primordial atmosphere is increasingly recognized as a key process in shaping planetary envelope compositions. This coupling should strongly influence gas accretion, yet its role during the disk-embedded stage remains poorly constrained.
We develop a time-dependent model that couples solid accretion, nebular-gas accretion, and water enrichment and partitioning through magma–atmosphere interactions, along with post-disk thermal evolution and escape.
We find that, for super-Earth-mass planets, water production is generally limited by the magma oxygen budget and typically ceases before disk dispersal. Subsequent nebular-gas accretion dilutes the envelope toward hydrogen-dominated compositions, largely independent of the initial magma redox state. 
This establishes an upper bound on the envelope water fraction---the oxygen exhaustion limit---primarily set by the reactive-oxygen inventory and the planet mass. After disk dispersal, degassing increases the water fraction only in Earth-mass planets undergoing strong escape, while super-Earths exhibit little change because surface pressures are hardly affected by escape.
Magma–atmosphere coupling alone therefore cannot maintain water-rich envelopes in sub-Neptunes and produces a strong mass–composition relation imposed by the oxygen-exhaustion limit. 
Highly enriched sub-Neptunes would therefore imply additional mechanisms such as late volatile delivery or post-disk giant impacts. The relation between planetary radius and envelope composition offers a means to infer magma properties, providing a pathway to connect present-day observables with early formation histories.
\end{abstract}



\section{Introduction}

Exoplanet surveys have shown that planets with sizes between Earth and Neptune are the most common outcome of planet formation~\citep[e.g.,][]{Howard2013,Petigura+2017,Zhu+2018}.  
These intermediate-sized planets span a wide range of masses and radii and are typically classified as super-Earths 
and sub-Neptunes. 
Their mass--radius relations indicate that many sub-Neptunes retain substantial H/He envelopes.

Such primordial H/He envelopes naturally arise because growing planets embedded in protoplanetary disks gravitationally capture nebular gas~\citep[e.g.,][]{Mizuno+1978,Ikoma+Genda2006,Bodenheimer+Lissauer2014,Lee+Chiang2015}.  
For super-Earth-mass planets, these envelopes can exceed 1\% of the total mass and strongly affect their observed radii~\citep{Lopez+Fortney2014}.  
Although they originate as H/He-dominated, primordial envelopes can be enriched in heavier volatiles, most notably water, C/O-bearing species, and even silicate vapor, which substantially modify their structure and evolution.  
Understanding how such enrichment operates during formation is therefore essential for linking present-day atmospheric compositions to planetary formation histories~\citep{Lichtenberg+etal2025}.

Various mechanisms have been proposed to enrich primordial envelopes.  
Exogenous pathways include the accretion of water-rich solids~\citep{Venturini+2016,Ormel+2021} and the capture of volatile-rich nebula gas produced by sublimating icy solids in the inner disk~\citep{Booth+2017,Schneider+Bitsch2021}.  
Endogenous enrichment arises from redox reactions between the envelope and the magma ocean, producing water by oxidation of hydrogen by FeO and other oxygen-bearing species~\citep{Ikoma+Genda2006,Kite+etal2020,Kite+Schaefer2021,Schlichting+Young2022,Seo+etal2024,Tian+Heng2024,2025ApJ...988L..55W}.  
In these cases, the equilibrium water abundance is set by the magma oxygen fugacity and the coupling dynamics between magma and envelope.

Recent laboratory studies support the plausibility of strong volatile enrichment during the formation stage.  
Experiments under conditions comparable to the magma–atmosphere interface show that water-rich equilibria are favored~\citep{Miozzi+etal2025}, and that even FeO-poor magmas can generate high water fractions by reducing silicates by metallic iron and hydrogen~\citep{Horn+etal2025}.  
These results suggest that substantial volatile enrichment of primordial atmospheres could be common and highlight the importance of understanding how these processes operate along with other planet-forming processes.

Water enrichment significantly affects the thermal structure and accretion of the envelope.  
By increasing the mean molecular weight and altering the adiabatic gradient through molecular dissociation and latent heat release, 
enriched envelopes can acquire much larger masses than H/He-dominated ones~\citep{Hori+Ikoma2011,Venturini+2015,Venturini+2016,Kimura+Ikoma2020,Kimura+Ikoma2022}.  
However, most previous studies that considered water enrichment during the disk-embedded phase neglected the dissolution of the water into the magma.  
In reality, water is highly soluble in silicate melt~\citep[e.g.,][]{Papale1997}, allowing a large fraction of volatile inventory to be stored in the interior, which can buffer atmospheric composition and modulate gas accretion efficiency.

Recent works have examined chemical interactions and volatile partitioning between atmospheres and magma oceans, mainly in the post-disk stage—when no further gas accretion occurs~\citep[e.g.,][]{Kite+etal2020,Lichtenberg+etal2021,Schlichting+Young2022,Bower+etal2022,Seo+etal2024}.  
Some studies explored interactions during solid accretion~\citep[e.g.,][]{Olson+Sharp2019}, but without accounting for feedback on atmospheric structure.  
Consequently, it remains unclear how primordial envelopes form and evolve when both redox-driven water production and disk-gas accretion operate simultaneously.

In this study, we develop a comprehensive, time-dependent model that extends previous post-disk frameworks into the disk-embedded phase.  
Our model self-consistently couples solid accretion, H/He gas accretion, redox-driven water enrichment, water dissolution into magma, and their feedback on envelope structure, and further follows thermal contraction and photoevaporative loss after disk dispersal.  
This enables us to quantify how the interplay among magma redox chemistry, volatile partitioning, and envelope accumulation governs the formation and final properties of super-Earths and sub-Neptunes.

\section{Model}

\begin{figure}
    \centering
    \includegraphics[width=\linewidth]{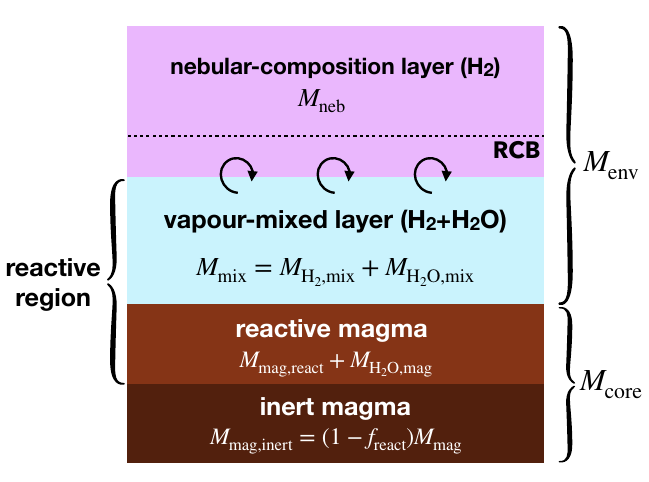}
    \caption{Schematic structure of the planet in our model. From top to bottom, the planet consists of four layers: a nebular-composition envelope (pure H$_2$), a vapor-mixed envelope (H$_2$ + H$_2$O), a reactive magma layer, and a non-reactive (inert) magma layer. Only the vapor-mixed envelope and the reactive magma are assumed to interact. If the radiative–convective boundary (RCB) lies within the nebular-composition layer, its convective part is assumed to mix with the vapor-mixed layer. See text for details.}
    \label{fig:planet_structure}
\end{figure}

We investigate how planets embedded in disk gas grow and accrete envelopes, while accounting for the enrichment of the envelope with water via interactions with magma and the partitioning of water between the envelope and magma. 
Our model self-consistently simulates (i) planetary growth by solid accretion, (ii) the formation and thermal evolution of the primordial envelope, (iii) water enrichment of the envelope through magma–gas interactions, and (iv) dissolution of the produced water into magma. 
We also follow the thermal evolution and atmospheric loss after disk dispersal, including redistribution of water between the envelope and magma. 
In this section, we describe the planetary structure assumed in the model and our treatment of envelope formation and evolution coupled with magma interaction.

\subsection{Planet structure}

We divide the planet, with total mass $M_{\rm p}$, into a ``core'' ($M_{\rm core}$) and an ``envelope'' ($M_{\rm env}$). 
The core is assumed to be fully molten and composed of rocky magma ($M_{\rm mag}$) together with dissolved H$_2$O 
($M_{\rm H_2O,mag}$). 
We further specify the fraction of magma that can chemically interact with the envelope, hereafter the ``reactive magma'', with mass
\begin{equation}
    M_{\rm mag,react} = f_{\rm react} M_{\rm mag},
\end{equation}
where $f_{\rm react}$ is a free parameter. 
We also denote by $M_{\rm O,react}$ the mass of oxygen atoms in the reactive magma that can participate in reactions with the envelope. 
This is not the total oxygen content of the magma, but only the portion available for redox reactions (e.g. O bound in FeO). 
We hereafter refer to such oxygen as ``reactive oxygen''. 
In our model, we do not specify the oxide species in magma; instead, we parameterise the amount of reactive oxygen. 
The quantity $M_{\rm O,react}$ increases through solid accretion and decreases as oxygen is consumed in water-producing reactions with the envelope (see below).

The envelope is divided into two layers: a lower ``vapor-mixed envelope'' and an upper ``nebular-composition envelope''. 
The vapor-mixed layer consists of H$_2$ and H$_2$O and is assumed to be in full equilibrium with the magma. 
In this layer, H$_2$ reacts with magma to form H$_2$O, and the water mass fraction, $X_{\rm H_2O,mix}$, is set equal to the equilibrium value $X_{\rm H_2O,eq}$. 
In reality, $X_{\rm H_2O,eq}$ is determined by the oxygen fugacity of the magma (i.e. its redox state). 
In our model, we take $X_{\rm H_2O,eq}$ as an input parameter.
For instance, magma buffered by the iron–w\"{u}stite equilibrium yields $X_{\rm H_2O,eq} \sim 0.8$--0.9. 
The water produced in the reactions is partitioned between the vapor-mixed envelope and the reactive magma, with the fraction residing in the envelope being uniformly distributed throughout the vapor-mixed layer. 
Thus, we always assume a uniform $X_{\rm H_2O,mix}$ in this layer.
\rev{
This assumption of instantaneous equilibration between the vapor-mixed envelope and the magma ocean is justified because both the envelope and the fully molten magma ocean are expected to be vigorously convective during the formation stage, with characteristic mixing timescales shorter than $\sim$1~yr~\citep{Kimura+Ikoma2020,Solomatov2000,Elkins-Tanton2008}.
These timescales are much shorter than the timestep used in our simulations, supporting the equilibrium approximation at each timestep.
}

If the reactive oxygen in the magma is fully consumed, water production ceases and $X_{\rm H_2O,mix}$ deviates from $X_{\rm H_2O,eq}$. 
In this case, its value is determined by the partitioning equilibrium with the reactive magma.

The upper envelope is chemically isolated from the magma and retains the nebular composition. 
We assume that this consists solely of hydrogen, neglecting minor species; for near-solar compositions, these components have a negligible effect on the envelope structure. 
The calculation of this two-layer envelope structure is described in \S~\ref{sec:model_phaseII}.

\subsection{Evolution of disk gas and planetary core}

The planetary core mass and radius, together with the disk gas pressure and temperature, define the boundary conditions for the envelope structure. 
Here we adopt a simplified model of disk gas dissipation and core growth.

Simulations of disk evolution including viscous diffusion and photoevaporation indicate that the disk gas density initially decreases on the viscous diffusion timescale. 
Once a gap opens due to efficient photoevaporation, the inner disk disperses rapidly on a timescale of $\sim 10^4$--$10^5$~yr~\citep[e.g.,][]{Clarke+2001,Alexander+2006a}. 
To approximate this behavior, we evolve the disk gas pressure $P_{\rm disk}$ as
\begin{align}
    P_{\rm disk}(t) &= P_{\rm disk,0}\exp(-t/t_{\rm disk}) \notag \\
    &\qquad \times
    \begin{cases}
        1 & \quad (t \le t_{\rm disk}) \\
        \exp\!\left[-\dfrac{t-t_{\rm disk}}{10^5~{\rm yr}}\right] & \quad (t > t_{\rm disk}), \\        
    \end{cases}
\end{align}
where the initial pressure is $P_{\rm disk,0}=1$~Pa and the dissipation timescale $t_{\rm disk}$ is an input parameter. 
Disk temperature $T_{\rm disk}$ is also treated as an input and remains constant throughout the simulation.

For core growth, we prescribe a solid accretion rate $\dot{M}_{\rm acc}$. 
The ``tentative'' magma mass, $M_{\rm mag}^0$, is then
\rev{
\begin{equation}
    M_{\rm mag}^0(t) = M_{\rm mag}(t_0) + \dot{M}_{\rm acc}\,\Delta t_{\rm sys},
\end{equation}
where $t_0$ and $\Delta t_{\rm sys}$ are the time at the previous step and the system timestep ($=t-t_0$), respectively.}
\rev{
Here the system timestep is set to be
\begin{equation}
    \Delta t_{\rm sys} = \min(0.05t, 10^4~{\rm yr}),
\end{equation}
which ensures that the timestep remains shorter than the characteristic growth timescales of both the core and the envelope, as well as the disk dissipation timescale.
}
Note that $M_{\rm mag}^0(t)$ does not represent the actual magma mass at time $t$, because the true mass is reduced by the amount of reactive oxygen consumed in water production during this step, which is determined after computing the envelope structure. 
Similarly, the tentative mass of reactive oxygen is
\begin{equation}
    M_{\rm O,react}^0(t) = M_{\rm O,react}(\rev{t_0}) + f_{\rm O,react}\,\dot{M}_{\rm acc}\,\Delta t_{\rm sys}.
\end{equation}
We assume a constant $\dot{M}_{\rm acc}$ until the total accreted mass reaches the isolation mass $M_{\rm iso}$, an input parameter, after which accretion ceases ($\dot{M}_{\rm acc}=0$). 
Specifically, we adopt $\dot{M}_{\rm acc} = 1\times 10^{-5} M_\oplus\,{\rm yr}^{-1}$ during the accretion phase. 
As shown in the results, most envelope properties are determined after solid accretion ends; hence the precise choice of $\dot{M}_{\rm acc}$ has little effect on the outcome. 

The tentative core mass is then given by
\begin{equation}
    M_{\rm core}^0(t) = M_{\rm mag}^0(t) + M_{\rm H_2O,mag}(\rev{t_0}),
\end{equation}
which is used to set the inner boundary conditions for the envelope structure.

\subsection{Envelope structure and water partitioning model}

The formation and evolution of the envelope are modelled following \cite{Kimura+Ikoma2022}. 
We divide the process into three phases (I–III):
\begin{itemize}
    \item \textbf{Phase I:} The envelope is in a hydrostatic and thermal steady state. Planets are in this phase during solid accretion.
    The entire envelope is assumed to be equilibrated with the magma and the water produced is uniformly mixed throughout. 
    Thus, only the vapor-mixed layer exists in this phase, with mass $M_{\rm mix}$.
    \item \textbf{Phase II:} After solid accretion ceases, the vapor-mixed envelope cools and contracts, enabling further accretion of disk gas. 
    A nebular-composition layer then develops on top of the vapor-mixed layer, with mass $M_{\rm neb}$. 
    Only the convective part of this layer is assumed to mix with the underlying vapor-mixed layer. 
    In this phase, we calculate the quasi-static contraction of the two-layer envelope.
    \item \textbf{Phase III:} After disk gas dispersal, the envelope undergoes a long-term thermal evolution and begins to lose mass through escape. 
\end{itemize}

In the following subsections, we describe the model used to calculate the envelope structure and water partitioning, and how each phase is simulated.

\subsection{Envelope structure and water partitioning}

We adopt the 1D internal structure model of \cite{Kimura+Ikoma2020}, solving the standard stellar structure equations:
\begin{align}
    \pdv{P}{R} &= -\frac{GM_R \rho}{R^2}, \label{eq:dPdr} \\
    \pdv{T}{R} &= -\frac{GM_R \rho}{R^2}\frac{T}{P}\nabla, \label{eq:dTdr} \\
    \pdv{M_R}{R} &= 4\pi R^2 \rho, \label{eq:dmdr}
\end{align}
where $P$, $T$, and $\rho$ are the pressure, temperature, and density of the envelope gas, $R$ is the radial distance from the planetary centre, and $M_R$ is the enclosed mass. 
The temperature gradient $\nabla \equiv \dv*{\log T}{\log P}$ is chosen according to radiative diffusion or convection (dry or moist adiabat).

The envelope is assumed to consist only of H and O. 
We use the hydrogen EOS of \cite{Chabrier+2021} and the water EOS of \cite{Haldemann+2020}. 
Latent heat release from water condensation is included in the adiabatic gradient. 
The opacities  are taken from the gas opacity table of \cite{Kimura+Ikoma2020} and the dust opacity of \cite{Semenov+2003}, with the dust depletion computed following \cite{Ormel2014}. 
See \cite{Kimura+Ikoma2020} for details.

The boundary conditions are set as follows: at the Bondi radius $R_{\rm B}$, $T=T_{\rm disk}$ and $P=P_{\rm disk}$; at the core radius $R_{\rm core}$, $M_R = M_{\rm core}^0$. 
The solid core radius is computed using \cite{Fortney+2007}:
\begin{align}
\begin{split}
    R_{\rm core} 
    &= (0.0592f_{\rm rock}+0.0975)(\log M_{\rm core}^0)^2 \\
    &\quad + (0.2337f_{\rm rock}+0.4938)\log M_{\rm core}^0 \\
    &\quad + (0.3102f_{\rm rock}+0.7932),
\end{split}
\label{eq:Rcore}
\end{align}
where $f_{\rm rock}$ is the Si/(Si+Fe) mass ratio, set to 0.66 (Earth-like). 
We neglect the effects of magma being molten and water dissolution, both of which can increase the core radius~\citep{bower_linking_2019,Dorn+Lichtenberg2021}. 
Although $M_{\rm core}^0$ is used as the boundary condition, the actual core mass depends on water dissolution and oxygen consumption, which are determined from the envelope calculation. 
Although iterative treatment would be thus more realistic, the associated corrections to core mass and radius are minor.

For a given water fraction $X_{\rm H_2O,mix}$ and constant envelope luminosity $L$, we compute the envelope structure. 
The method for evaluating $L$ depends on the evolutionary phase (see below).

The mass fraction of water dissolved in the magma is calculated using the solubility law:
\begin{equation}
    X_{\rm H_2O,mag} = \frac{M_{\rm H_2O,mag}}{M_{\rm mag,react}+M_{\rm H_2O,mag}} 
    = \alpha P_{\rm H_2O,surf}^{1/\beta},
    \label{eq:xH2O_r}
\end{equation}
where $P_{\rm H_2O,surf}$ is the partial pressure of H$_2$O at the magma surface, given by the ideal gas relation:
\begin{equation}
    P_{\rm H_2O,surf} = \frac{\mu}{\mu_{\rm H_2O}} X_{\rm H_2O,mix} P_{\rm surf},
\end{equation}
with $\mu$ the mean molecular weight at the bottom of the envelope. 
We adopt the solubility coefficients $\alpha = 215~{\rm ppmw\,bar^{-1/\beta}}$ and $\beta = 1/0.7$ from the fit in \cite{Bower+etal2022}, based on the experiments of basalt rock \citep{wilson_ascent_1981}. 
Although these data cover only $\sim$1–6~kbar, the surface pressure in our simulations often exceeds 10~kbar, and the surface temperatures can reach $>5000$~K. 
However, we still adopt this law as an approximation because the behaviour of H$_2$O–magma systems under such extreme conditions is poorly constrained.
We should also note that we neglect H$_2$ dissolution into magma, as it is orders of magnitude less efficient than H$_2$O dissolution.
These treatments correspond to the lower limit of volatile dissolution, because the water and hydrogen become fully miscible with magma above certain points of pressure and temperature~\citep[e.g.,][]{bureau_complete_1999,mibe_second_2007,gao_phase_2024,Young+etal2024,Stixrude+Gilmore2025a}.

The total water mass in the planet is then
\begin{equation}
    M_{\rm H_2O,tot} = X_{\rm H_2O,mix}M_{\rm mix} + M_{\rm H_2O,mag}.
\end{equation}
The remaining reactive oxygen mass in the magma is obtained from the change in $M_{\rm H_2O,tot}$, since all oxygen in water originates from the magma:
\begin{equation}
    M_{\rm O,react}(t) = M_{\rm O,react}^0(t) - \frac{8}{9}\Delta M_{\rm H_2O,tot},
\end{equation}
where
\begin{equation}
    \Delta M_{\rm H_2O,tot} = M_{\rm H_2O,tot}(t) - M_{\rm H_2O,tot}(\rev{t_0}).
\end{equation}

\subsection{Phase I: Purely hydrostatic phase}

In Phase I, the envelope structure and water partitioning are calculated assuming the luminosity $L$ arises solely from the solid core:
\begin{equation}
    L_{\rm core} = L_{\rm acc} + L_{\rm radio},
    \label{eq:Lcore}
\end{equation}
where $L_{\rm acc}$ and $L_{\rm radio}$ are the luminosities due to solid accretion and radioactive decay, respectively. 
The accretion luminosity is given by
\begin{equation}
    L_{\rm acc} = \frac{G M_{\rm core}^0 \dot{M}_{\rm acc}}{R_{\rm core}},
    \label{eq:Lacc}
\end{equation}
and the radiogenic luminosity is set to $L_{\rm radio} = 2\times 10^{20}(M_{\rm core}^0/M_\oplus)~{\rm erg~s^{-1}}$~\citep{Guillot+1995}.

If $M_{\rm O,react}^0(t) > 0$, the envelope is in complete equilibrium with the magma, and we set $X_{\rm H_2O,mix} = X_{\rm H_2O,eq}$ to calculate the envelope structure and water partitioning. 
If $M_{\rm O,react}(t)$ becomes zero or if $M_{\rm O,react}^0(t) = 0$, no further water is produced. 
The total water mass is then fixed as
\begin{equation}
    M_{\rm H_2O,tot}(t) = M_{\rm H_2O,tot}(\rev{t_0}) + \frac{9}{8}M_{\rm O,react}^0(t).
    \label{eq:MH2O_tot}
\end{equation}
The envelope calculations are iterated with the updated $X_{\rm H_2O,mix}$ until the resulting $M_{\rm H_2O,tot}$ matches Eq.~\eqref{eq:MH2O_tot}.

\subsection{Phase II: Quasi-static thermal evolution}
\label{sec:model_phaseII}

After solid accretion ceases, the envelope contracts quasi-statically and disk gas accumulates atop the vapor-mixed layer, forming a nebular-composition layer. 
Only the convective portion of this layer is assumed to mix with the underlying vapor-mixed envelope. 
For numerical simplicity, we assume that H$_2$ in the convective region from the previous timestep is mixed into the vapor-mixed layer in the current timestep, enabling reaction with magma.

\rev{
The thermal evolution of the envelope is computed using total energy conservation, following \cite{Kimura+Ikoma2022} and previous gas-giant formation studies~\citep{Papaloizou+Nelson2005,Mordasini+2012b,Fortier+2013,Piso+Youdin2014,Venturini+2016}.
In this approach, we first assume the envelope luminosity $L$ and compute the corresponding internal structure, and then derive the timestep $\Delta t$ required to evolve from the previous state at $t_0$ to this structure.
Thus, the timestep $\Delta t$ is an outcome of the calculation and does not necessarily coincide with the system timestep $\Delta t_{\rm sys}$.
We therefore denote the time obtained from this procedure as $t'$, such that $\Delta t = t' - t_0$.
}

First, to determine ($X_{\rm H_2O,mix}$, $M_{\rm mix}$, $M_{\rm neb}$) for a given luminosity $L$, we proceed as follows. 
If $M_{\rm O,react}^0(\rev{t'}) > 0$, the vapor-mixed layer remains in equilibrium with magma, and we set $X_{\rm H_2O,mix} = X_{\rm H_2O,eq}$. 
The total hydrogen (H) mass in the vapor-mixed layer and reactive magma changes only via mixing with the nebular-composition envelope:
\begin{align}
    M_{\rm H}(\rev{t'}) &= M_{\rm H_2,mix}(\rev{t_0}) + \frac{1}{9}M_{\rm H_2O,tot}(\rev{t_0}) \notag \\
    & \qquad + M_{\rm neb,conv}(\rev{t_0}),
    \label{eq:MH_tot}
\end{align}
where $M_{\rm H_2,mix}$ is the H$_2$ mass in the vapor-mixed envelope and $M_{\rm neb,conv}$ is the convective mass of the nebular-composition layer. 
We iterate $M_{\rm mix}$ in the envelope calculation until Eq.~\eqref{eq:MH_tot} is satisfied and then derive $M_{\rm neb}$ from the self-consistent solution.

If $M_{\rm O,react}^0(\rev{t'}) = 0$, no further water is produced and the H$_2$ mass evolves exclusively through mixing:
\begin{equation}
    M_{\rm H_2,mix}(\rev{t'}) = M_{\rm H_2,mix}(\rev{t_0}) + M_{\rm H_2,neb,conv}(\rev{t_0}).
    \label{eq:MH2_tot}
\end{equation}
The relation between $X_{\rm H_2O,mix}$ and $M_{\rm mix}$ is then
\begin{equation}
    M_{\rm mix} = \frac{M_{\rm H_2,mix}}{1 - X_{\rm H_2O,mix}},
\end{equation}
and the envelope calculations are iterated to satisfy Eq.~\eqref{eq:MH2_tot}.

\rev{The envelope luminosity $L$ is given by the energy conservation equation}:
\begin{align}
    L(\rev{t'}) &= L_{\rm core} -\frac{E_{\rm env}(\rev{t'})-E_{\rm env}(\rev{t_0})}{\Delta t}  \notag \\
    & \qquad + e_{\rm gas} \frac{M_{\rm env}(\rev{t'}) - M_{\rm env}(\rev{t_0})}{\Delta t} ,
    \label{eq:energy_conserv}
\end{align}
where $E_{\rm env}$ is the total energy of the envelope,
\begin{equation}
    E_{\rm env} = \int_{M_{\rm core}^0}^{M_{\rm p}} \Big(u - \frac{GM_R}{R}\Big) \dd{M_R},
\end{equation}
$u$ is the specific internal energy and $e_{\rm gas}$ is the energy per unit mass of the disk gas at the outer boundary.

The core luminosity is expressed as
\begin{equation}
    L_{\rm core} = L_{\rm cool} + L_{\rm radio},
    \label{eq:Lcore_II}
\end{equation}
with the cooling term
\begin{equation}
    L_{\rm cool} = -M_{\rm core}^0 C_{\rm rock} \frac{T_{\rm surf}(\rev{t'}) - T_{\rm surf}(\rev{t_0})}{\Delta t},
    \label{eq:Lcool}
\end{equation}
where $C_{\rm rock} = 1.2\times 10^7~{\rm erg~g^{-1} K^{-1}}$ is the specific heat of the rock and $T_{\rm surf}$ is the temperature at the magma surface. 

Substituting Eqs.~\eqref{eq:Lcore_II} and \eqref{eq:Lcool} into Eq.~\eqref{eq:energy_conserv} gives
\begin{equation}
    \Delta t = \frac{\Delta E}{L_{\rm acc} + L_{\rm radio} - L},
    \label{eq:energy_conserv2}
\end{equation}
with
\begin{align}
    \Delta E &= E_{\rm env}(\rev{t'}) - E_{\rm env}(\rev{t_0}) \notag \\
    & \quad + M_{\rm core} C_{\rm rock} \left[T_{\rm surf}(\rev{t'}) - T_{\rm surf}(\rev{t_0})\right] \notag \\
    &\quad - e_{\rm gas} \left[M_{\rm env}(\rev{t'}) - M_{\rm env}(\rev{t_0})\right].
\end{align}

Thus, we first assume \rev{$L(t')$} and integrate the internal structure equations to compute $E_{\rm env}(t')$ and $M_{\rm env}(t')$, then determine $\Delta t$ from Eq.~\eqref{eq:energy_conserv2}. 
\rev{
If $\Delta t > \Delta t_{\rm sys}$, $L(\rev{t'})$ is iteratively adjusted until $\Delta t = \Delta t_{\rm sys}$.
If $\Delta t < \Delta t_{\rm sys}$, we advance the time to $t_0 = t'$ and repeat the procedure until $t'$ reaches the system time $t$.
}

\subsection{Phase III: Post-disk Thermal Evolution and Loss} 
\label{sec:model_phaseIII}

After disk dispersal, the planetary radius and envelope mass evolve due to thermal contraction and photoevaporation (Phase III). 
We assume that the envelope detaches from the disk gas when $P_{\rm disk}/P_{\rm disk,0} < 10^{-5}$, which roughly corresponds to a radial optical depth for stellar XUV below unity at $\lesssim 1$~au.

\subsubsection{Thermal evolution of the envelope}

The thermal evolution of the envelope is calculated as in \S~\ref{sec:model_phaseII} to determine the planetary radius $R_{\rm p}$ and the corresponding water partitioning state. 
Here, $R_{\rm p}$ is defined at a pressure level of 10~mbar.

For numerical convenience, the upper part of the envelope (called ``atmosphere'' hereafter) is treated separately from the deeper part (the envelope), \rev{
following standard thermal evolution models for giant planets and sub-Neptunes~\citep[e.g.,][]{Fortney+2007,Kurosaki+Ikoma2017}.
The boundary between these two regions is placed at the radius where the optical depth to stellar visible radiation equals 10, such that stellar irradiation has a negligible impact on the thermal structure of the deep envelope.
} 
The outer boundary conditions for the \rev{deep} envelope are
\begin{equation}
    P = P_{\rm out}, \quad T = T_{\rm out} \qquad {\rm at}\, M_R = M_{\rm p}.
\end{equation}
Here $P_{\rm out}$ and $T_{\rm out}$ are obtained by computing the radiative-convective structure of the \rev{upper} atmosphere following \cite{Kimura+Ikoma2022}.
For simplicity, we set the equilibrium temperature $T_{\rm eq}$ equal to the disk temperature $T_{\rm disk}$.

We neglect the effects of magma solidification, which would enhance H$_2$O degassing. 
In most of our simulations, the magma ocean remains molten, so the water partitioning is primarily affected by the decreasing pressure at the magma surface caused by envelope escape.
Long-lived molten magma in sub-Neptunes has also been suggested in previous studies~\citep{Vazan+etal2018b,nicholls_convective_2024,Calder+etal2025}.

\begin{table}[t]
    \centering
    \caption{Parameters and nominal values in our simulations}
    \label{tab:parameter}    
    {
    \renewcommand{\arraystretch}{1.4}
    \begin{tabular}{lll} \toprule
     Symbol & Meaning & Nominal Value \\ \midrule
     $t_{\rm disk}$  & disk dissipation timescale & $2\times 10^{6}$~yr  \\
     $T_{\rm disk}$ & disk temperature & 500~K \\
     $f_{\rm react}$ & Reactive magma fraction & 1.0 \\
     \shortstack[l]{{} \\ $f_{\rm O,react}$ \\ {}} & 
     \shortstack[l]{{} \\ Reactive O fraction \\ in accreting solids}
     & \shortstack[l]{{} \\ 0.1 \\ {}} \\
     $\dot{M}_{\rm acc}$ & Solid accretion rate & $1\times 10^{-5}~M_\oplus/{\rm yr}$ \\     
     $M_{\rm iso}$ & Isolation mass & $3~M_\oplus$ \\
     \shortstack[l]{ {} \\ $X_{\rm H_2O,eq}$ \\ {} \\ {}} &
     \shortstack[l]{ {} \\
     H$_2$O mass fraction \\ in vapor-mixed envelope \\ at magma equilibrium }
     & \shortstack[l]{ {} \\ 0.5 \\ {} \\ {} \\ {} }\\
     \bottomrule
    \end{tabular}
    }
\end{table}

\begin{figure*}
    \centering
    \includegraphics[width=\linewidth]{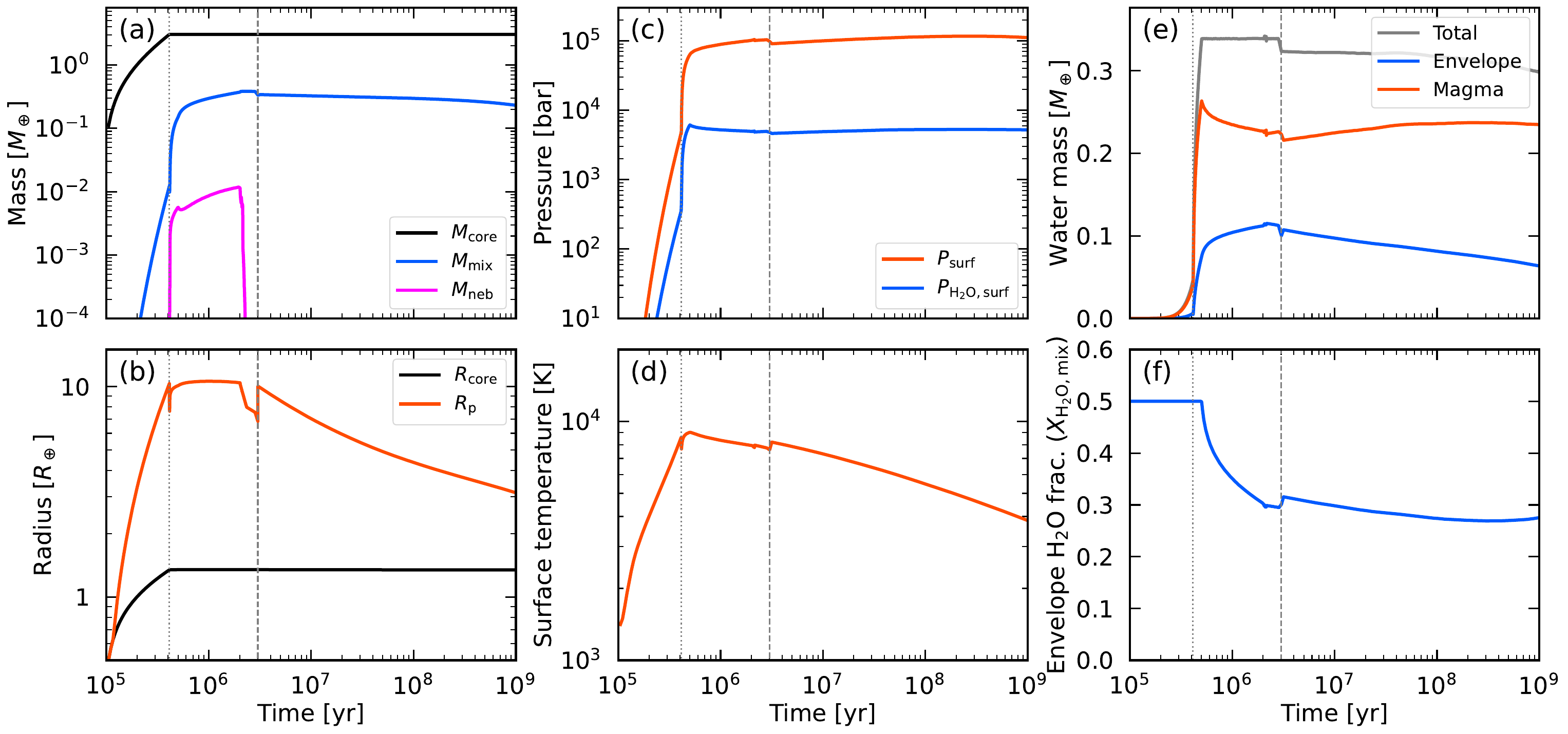}
    \caption{Time evolution of planetary properties for the nominal case (see Table~\ref{tab:parameter} for parameter values).
    Panel (a): Masses of the core (black), vapor-mixed envelope (blue), and nebular-composition envelope (magenta).
    Panel (b): Total planetary radius at 10~mbar (red) and core radius (black).
    Panel (c): Total (red) and H$_2$O partial (blue) pressures at the magma surface.
    Panel (d): Temperature at the bottom of the envelope.
    Panel (e): Water mass in the bulk planet (grey), magma (red), and vapor-mixed envelope (blue).
    Panel (f): Water mass fraction in the vapor-mixed envelope.
    Grey dotted and dashed lines indicate the termination of solid accretion and the time of disk dispersal, respectively.}
    \label{fig:MRPTWX_nominal}
\end{figure*}
\subsubsection{Envelope photoevaporation}

Envelope escape occurs due to stellar high-energy irradiation. 
%
%
Importantly, for the vapor-mixed envelope, hydrodynamic escape is less efficient due to enhanced radiative cooling from H$_2$O and related species~\citep{Yoshida+2022}. 
To take this effect into account, 
here we adopt the fitting formula of \cite{Yoshida+Gaidos2025}: 
\begin{equation}
    \dot{M}_{\rm esc} = \dot{M}_{\rm ref} 
    \qty(
    \frac{F_{\rm EUV}}{10^3F_{\rm EUV}^\oplus}
    )^a
    \qty(\frac{g}{4~{\rm m/s^2}})^{-3/2} \qty(\frac{M_{\rm p}}{5M_\oplus})^{1/2},    
\end{equation}
where $F_{\rm EUV}$ and $F_{\rm EUV}^\oplus$ are the stellar EUV fluxes at the planet and at Earth, respectively.
Here $F_{\rm EUV}$ is set to be $1\times 10^{-4}$ times the bolometric stellar insolation, which is calculated \rev{as $4\sigma T_{\rm eq}^4$, with $\sigma$ being the Stefan-Boltzmann constant
}.
The fitting parameters $\dot{M}_{\rm ref}$ and $a$ depend on the H$_2$O/H$_2$ number ratio $r_{\rm H_2O}$ and $F_{\rm EUV}$ (see \citealt{Yoshida+Gaidos2025}). 
We limit $r_{\rm H_2O} \le 0.1$ as the higher values are outside the fitted range, where radiative cooling saturates due to IR optical thickness.

We neglect the fractionation between H$_2$ and H$_2$O during escape, assuming that the vapor-mixed envelope retains its composition $X_{\rm H_2O,mix}$. 
While this approximation is valid for highly irradiated close-in planets, fractionation may affect the atmospheric lifetime for temperate planets~\citep{Yoshida+Gaidos2025}.

\subsection{Numerical settings}

Table~\ref{tab:parameter} summarises the input parameters and their nominal values. 
The adopted disk dissipation timescale represents typical lifetimes inferred from disk observations~\citep[e.g.,][]{Mamajek2009,Ansdell+2017,Richert+2018}, and the disk temperature is chosen as a representative value for close-in region where exogenous ice delivery is inefficient.  
For magma properties, we assume $f_{\rm react}=1$ as a nominal case corresponding to a fully molten, vigorously convecting magma in which the entire core can interact with the envelope.  
The value $f_{\rm O,react}=0.1$ roughly corresponds to the limiting case for an Earth-like bulk composition in which most iron is initially present as FeO.  
The equilibrium water mass fraction in the envelope, $X_{\rm H_2O,eq}=0.5$, corresponds to redox equilibria expected for moderately reducing magmas~\citep{Ikoma+Genda2006,Kite+etal2020,Seo+etal2024}. 
The initial planetary mass is set to $0.1~M_\oplus$, with a maximum integration time of $10^9$~yr. 
Simulations are terminated if the envelope mass exceeds the core mass, as runaway gas accretion would begin beyond this point, which is beyond the scope of this study. 

\section{Results}
\subsection{Nominal Case}

\begin{figure}
    \centering
    \includegraphics[width=\linewidth]{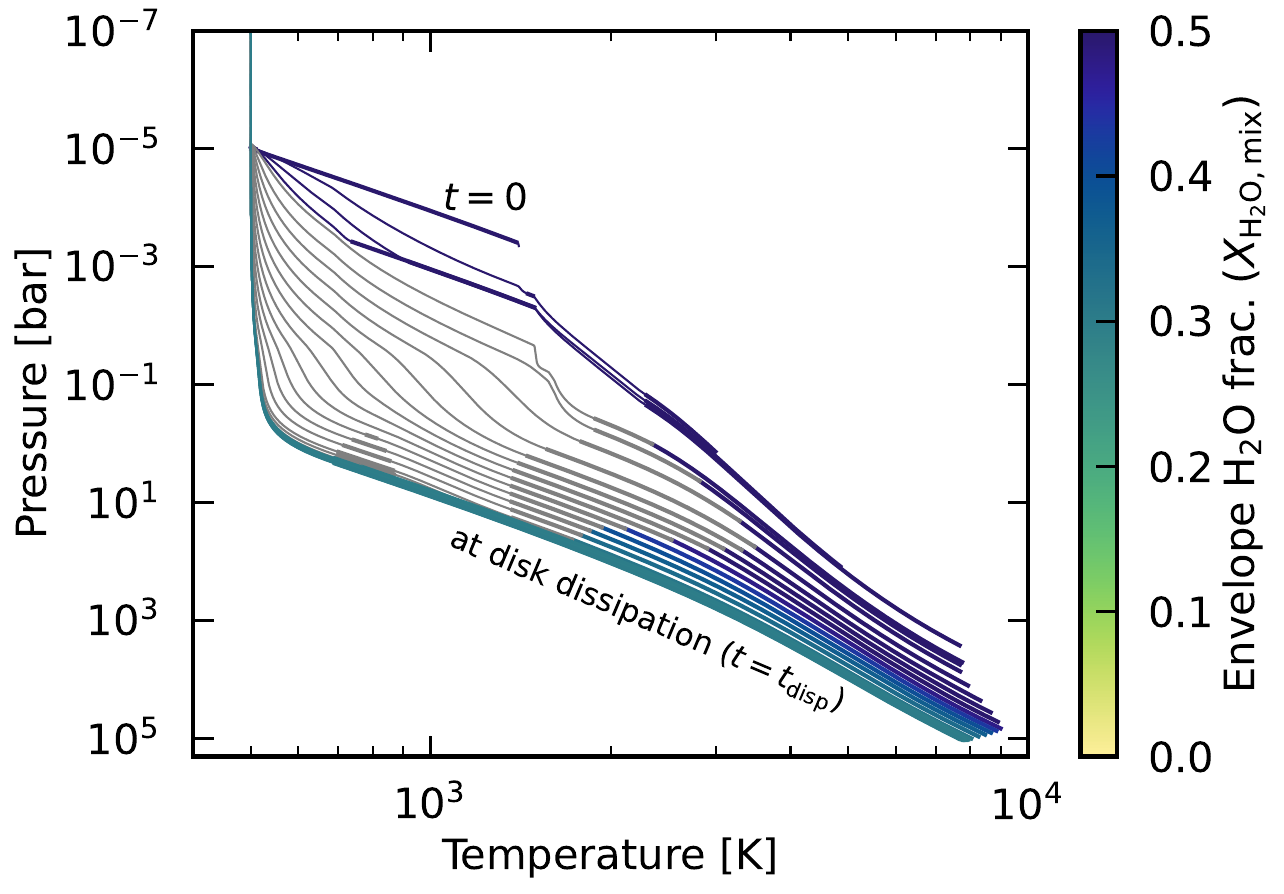}
    \caption{
    Time evolution of the envelope's pressure–temperature structure in the nominal case 
    (see also Fig.~\ref{fig:MRPTWX_nominal}).  
    The grey region shows the nebular-composition layer and the colored region shows the vapor-mixed layer, color-coded by its H$_2$O mass fraction ($X_{\rm H_2O,mix}$).  
    Thin and thick segments indicate radiative and convective regions, respectively.  
    The outer boundary is fixed at 500~K, while the pressure decreases from 1~Pa as the disk dissipates. 
    }
    \label{fig:PT_structure}
\end{figure}

We first present the result for the nominal case (Table~\ref{tab:parameter}) to illustrate how planets grow and how the interplay between the gas accretion and magma-envelope interactions shapes the resulting envelope properties in our model.
Figure~\ref{fig:MRPTWX_nominal} shows the time evolution of the planetary envelope and the distribution of water between the envelope and magma, and Fig.~\ref{fig:PT_structure} shows the time evolution of the pressure-temperature structure during the disk-embedded phase (Phase I \& II) in this case.

The planet reaches $\sim 3 M_\oplus$ within the first $\sim 4\times10^5$~yr.  
During the solid accretion phase (Phase~I), the vapor-mixed envelope grows as the core mass increases, and its deep region becomes convective with pressures $\gtrsim$~kbar and temperatures of nearly $10^4$~K at the time of isolation (Fig.~\ref{fig:PT_structure}).  
By this stage, the envelope mass already approaches $\sim$1\% of the planet mass.  
Once solid accretion ceases, envelope cooling and contraction accelerate nebular gas accretion (Fig.~\ref{fig:MRPTWX_nominal}(a)), forming a nebular-composition layer (shown in grey lines in Fig.~\ref{fig:PT_structure}).
As this accumulation proceeds and the nebular-composition layer becomes thicker, the deep part of this layer becomes convective (Fig.~\ref{fig:PT_structure}) and efficiently mixes with the underlying vapor-mixed layer.  
Throughout Phase~II, except immediately before disk dispersal, the interface between the mixed and nebular-composition layers lies near $\sim100$~bar, with temperature gradually decreasing with time.

Water production and partitioning evolve together with this structural evolution.  
Mixing of the nebular-composition layer into the vapor-mixed layer supplies additional hydrogen to react with the magma, significantly enhancing the total water mass in the planet (Fig.~\ref{fig:MRPTWX_nominal} (e)).  
Due to the high pressure at the magma surface, most of the produced water is dissolved into the magma rather than remaining in the envelope.

Because the oxidation reactions occur significantly just after the end of solid accretion, the reactive oxygen in the magma is quickly exhausted, limiting the water production and enrichment of the envelope. We term this limit the ``oxygen exhaustion limit''.
After that, the total water mass remains nearly constant until disk dispersal.  
However, as nebular gas continues to accrete, the vapor-mixed envelope is gradually diluted, causing the water mass fraction $X_{\rm H_2O,mix}$ to decrease with time (Figs.~\ref{fig:MRPTWX_nominal}(f) and \ref{fig:PT_structure}). 
Consequently, the H$_2$O partial pressure at the magma surface slightly decreases, whereas the total pressure continues to increase (Fig.~\ref{fig:MRPTWX_nominal} (c)).  
This drives partial degassing from the magma, slightly increasing the water mass in the envelope despite the overall decline in the water mass fraction.

After $\sim2$~Myr, the disk gas begins to dissipate rapidly, eroding the nebular-composition layer.  
The envelope then becomes single-layered again, fully composed of the vapor-mixed component.  
A fraction of this layer also escapes during the final stage of disk dispersal.  
At $\sim3$~Myr, once the disk gas is completely dissipated, photoevaporative mass loss begins, although its effect remains modest in this case.  
The discontinuities visible in Fig.~\ref{fig:MRPTWX_nominal} at this time are caused by the abrupt change in the outer boundary condition of the envelope model.

During the early post-disk evolution ($\lesssim0.1$~Gyr), envelope contraction increases the surface pressure, enhancing H$_2$O dissolution into the magma and slightly reducing the envelope water fraction.  
Later, as escape becomes dominant, the surface pressure decreases, triggering a partial degassing that slightly increases $X_{\rm H_2O,mix}$, although the total water mass still decreases.

Overall, the envelope composition and water partitioning are largely set during the disk-embedded phases, where gas accretion and oxygen exhaustion determine the water inventory.
Subsequent thermal evolution and escape only modestly modify the composition.  
In the nominal case, the planet ends with $M_{\rm p}\simeq 3M_\oplus$, $R_{\rm p}\simeq 3R_\oplus$, and an envelope comprising $\sim$10~wt.\% of the total mass, with a water mass fraction of $\sim$0.3 in the vapor-mixed layer.

\subsection{Effects of Initial Envelope Composition}

\begin{figure}
    \centering
    \includegraphics[width=\linewidth]{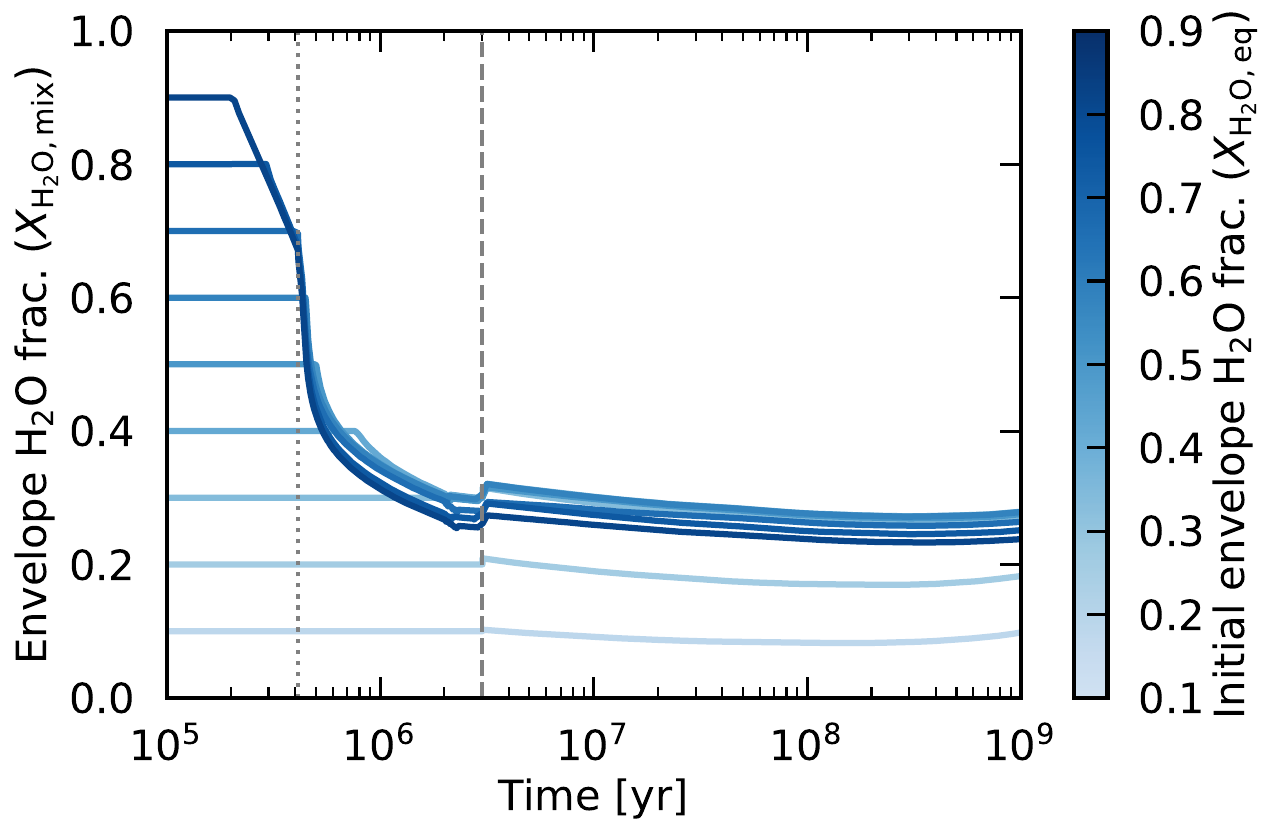}
    \caption{Time evolution of the water mass fraction in the vapor-mixed envelope ($X_{\rm H_2O,mix}$) for different initial values of $X_{\rm H_2O,eq}$. 
    Other parameters are identical to those listed in Table~\ref{tab:parameter}.
    Grey dotted and dashed lines indicate the termination of solid accretion and the time of disk dispersal, respectively.}
    \label{fig:tX_Xeq}
\end{figure}

Since the envelope mass strongly depends on its composition, we examine how the initial water mass fraction of the envelope, $X_{\rm H_2O,eq}$, affects the final envelope properties.  
Figure~\ref{fig:tX_Xeq} shows the time evolution of $X_{\rm H_2O,mix}$ for cases with $X_{\rm H_2O,eq}$ varying from 0.1 to 0.9 in increments of 0.1, while all other parameters are fixed to the nominal values (Table~\ref{tab:parameter}).

Despite the large variation in the initial composition, the resulting envelope compositions at both disk dispersal and the final evolutionary stage converge to a relatively narrow range.  
In particular, for $X_{\rm H_2O,eq} \ge 0.3$, the final water fraction stabilises at $X_{\rm H_2O,mix} \simeq 0.3$.  
This indicates that maintaining a highly water-enriched envelope throughout the formation phase (i.e. while the planet remains embedded in the disk) is extremely difficult.

This convergence arises from the ``oxygen exhaustion limit'' as explained in the previous section: when $X_{\rm H_2O,mix} \gtrsim 0.3$, the reactive oxygen in the magma becomes exhausted at some point during the embedded phase, and the water production does not occur anymore.
Before oxygen exhaustion, where reactive oxygen remains abundant and the envelope composition is fixed at $X_{\rm H_2O,mix} = X_{\rm H_2O,eq}$, planets with higher $X_{\rm H_2O,eq}$ accrete the nebular gas more efficiently~\citep{Kimura+Ikoma2020}.  
Consequently, reactive oxygen consumption proceeds more rapidly in high-$X_{\rm H_2O,eq}$ cases, leading to earlier oxygen exhaustion.  
For $X_{\rm H_2O,eq} \ge 0.7$, we find that this exhaustion limit occurs even before the termination of solid accretion.
In such cases, $X_{\rm H_2O,mix}$ starts decreasing already during solid accretion, as the newly accreted oxides are immediately reduced by the abundant hydrogen and converted into water.

After the reactive oxygen is depleted, further accretion of nebular gas dilutes the vapor-mixed envelope, causing $X_{\rm H_2O,mix}$ to decline with time.  
Since the enriched envelope quickly contracts after the termination of solid accretion, the gas accretion rate is regulated by the Kelvin-Helmholtz contraction of the outer nebular-composition layer.  
As a result, the accretion rate and subsequent evolution become largely independent of $X_{\rm H_2O,eq}$, and all models follow a similar evolutionary track as long as the oxygen exhaustion limit is reached.

Therefore, the envelope composition at disk dispersal is mainly determined by the total water mass (or equivalently, the total amount of reactive oxygen \rev{in magma}) and the envelope mass of the planet, rather than by the initial $X_{\rm H_2O,eq}$ that reflects the redox state of the magma.
\rev{
In other words, magmas with different redox states can yield similar final envelope compositions, provided that the total mass of reactive oxygen available for water production is the same.
}

\subsection{Effects of Planet Mass}

\begin{figure}
    \centering
    \includegraphics[width=\linewidth]{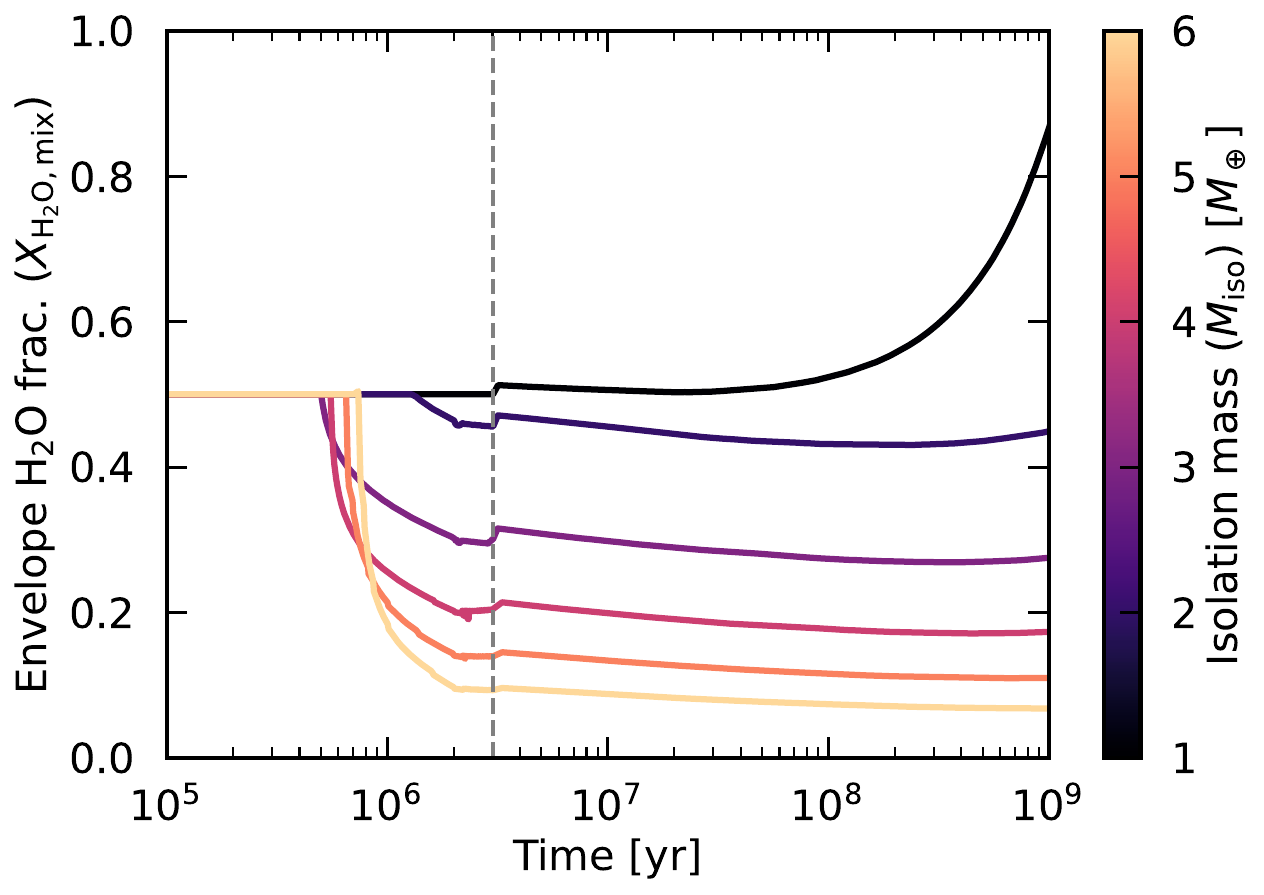}
    \caption{Time evolution of the water mass fraction in the vapor-mixed envelope ($X_{\rm H_2O,mix}$) for different isolation masses $M_{\rm iso}$, ranging from $1M_\oplus$ to $6M_\oplus$ in $1M_\oplus$ increments. 
    Other parameters are the same as in Table~\ref{tab:parameter}. 
    The grey dashed line marks the time of disk dispersal.}
    \label{fig:tX_Miso}
\end{figure}

\begin{figure*}
\centering
\includegraphics[width=0.8\linewidth]{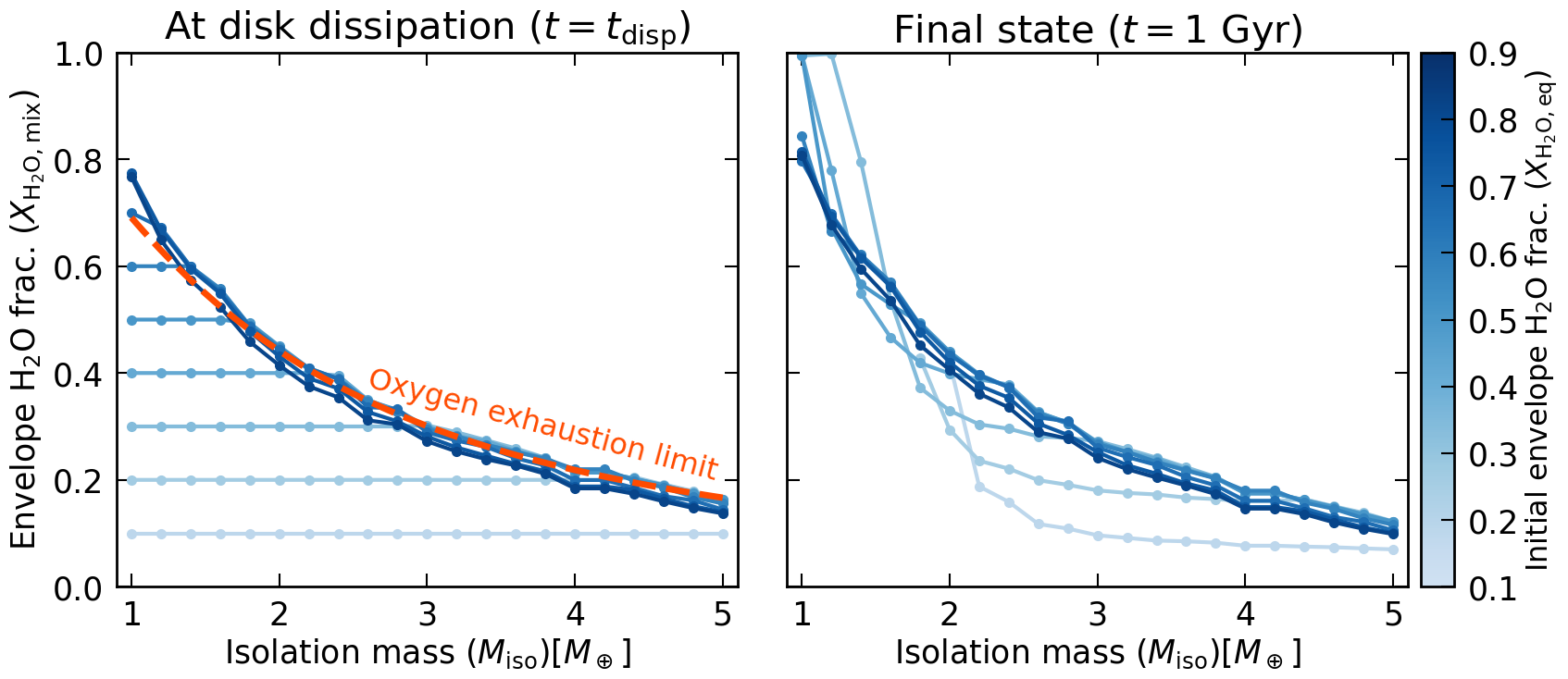}
\caption{Water mass fraction in the vapor-mixed envelope ($X_{\rm H_2O,mix}$) at the time of disk dispersal (left panel) and at the final state (1~Gyr; right panel) as a function of planetary isolation mass ($M_{\rm iso}$).
The colour of the lines indicates the initial equilibrium water mass fraction in the vapor-mixed envelope ($X_{\rm H_2O,eq}$).
All other parameters are fixed to the nominal values (Table~\ref{tab:parameter}).
The red dashed line in the left panel shows the semi-analytical estimate of the oxygen exhaustion limit (Eq.~\ref{eq:O_exhaustion_line}).
In the right panel, planets that have completely lost their primordial envelopes are excluded.}
\label{fig:Miso_Xenv_nominal}
\end{figure*}

Planetary mass strongly influences both the amount of accreted gas and the inventory of reactive oxygen available for water production.  
Figure~\ref{fig:tX_Miso} shows the time evolution of the envelope water fraction $X_{\rm H_2O,mix}$ for planets with isolation masses $M_{\rm iso}$ between $1M_\oplus$ and $6M_\oplus$, while all other parameters are fixed to the nominal values (Table~\ref{tab:parameter}).

We find that $X_{\rm H_2O,mix}$, both at disk dispersal and at the final stage, decreases with increasing planet mass.  
In particular, only the $1M_\oplus$ case exhibits a large increase in $X_{\rm H_2O,mix}$ after disk dispersal, whereas for $M_{\rm iso} \ge 2M_\oplus$, the envelope composition remains nearly constant during the post-disk phase.

The lower water fractions in more massive planets arise because the amount of accreted hydrogen increases more steeply with planet mass than the supply of reactive oxygen, which scales linearly with core mass.  
Although larger planets have more reactive oxygen overall, the more efficient gas accretion significantly dilute the vapor-mixed envelope once the oxygen exhaustion limit is reaced.
In contrast, lower-mass planets accrete nebular gas less efficiently, allowing them to maintain highly enriched envelopes for longer period.  
In our simulations, only the $1M_\oplus$ planet retains $X_{\rm H_2O,mix} = X_{\rm H_2O,eq}$ until disk dispersal, as the exhaustion of reactive oxygen does not occur.

The post-disk increase in $X_{\rm H_2O,mix}$ for the $1M_\oplus$ planet is driven by efficient envelope escape.  
At disk dispersal, the envelope mass is $\sim 0.01M_\oplus$, which decreases to $\sim 0.001M_\oplus$ by the end of the simulation due to photoevaporation.  
The resulting depressurization on the magma surface promotes the degassing of H$_2$O from the interior, increasing the relative water abundance in the remaining envelope.
Note that in this case, the temperature on the magma surface decreases to $\sim 1500$~ K in the final state due to the effective cooling of the core, resulting from the relatively thin envelope. Thus, the solidification of the inner part of the core would affect the water partitioning state, which is neglected in our model.
Since magma solidification generally results in efficient degassing of dissolved water, in reality the planet would end up having a thicker steam atmosphere.

In summary, even when the magma initially contains a substantial oxygen reservoir, the oxygen exhaustion commonly occurs in super-Earth-mass planets.  
Once this happens, it becomes difficult for such planets to retain highly water-enriched envelopes until the end of the disk phase.

\subsection{Final Envelope Composition}

Figure~\ref{fig:Miso_Xenv_nominal} summarizes the dependence of the water mass fraction in the vapor-mixed envelope, $X_{\rm H_2O,mix}$, on the planetary isolation mass $M_{\rm iso}$, both at the time of disk dispersal ($t_{\rm disp}$; left) and in the final state (1~Gyr; right).  
Simulations were conducted for $M_{\rm iso}$ between $1M_\oplus$ and $5M_\oplus$ and for equilibrium envelope compositions $X_{\rm H_2O,eq}$ ranging from 0.1 to 0.9, with all other parameters fixed to the nominal values (Table~\ref{tab:parameter}).

\medskip
\noindent
\textbf{At the time of disk dispersal}\\
Irrespective of the assumed equilibrium composition $X_{\rm H_2O,eq}$ (which reflects the magma redox state), planets that grow to super-Earth masses exhibit relatively low $X_{\rm H_2O,mix}$ at $t_{\rm disp}$.  
A distinct upper boundary appears in $X_{\rm H_2O,mix}$ for each $M_{\rm iso}$, which decreases systematically with increasing planet mass.  
As shown in the previous sections, this behavior arises from the oxygen exhaustion limit.
The key point of this feature is that, even when the redox reaction leads to much higher $X_{\rm H_2O,eq}$ than this limit, the envelope composition converges toward this limit by the end of the disk phase.
In contrast, when $X_{\rm H_2O,eq}$ is below the limit, the envelope maintains $X_{\rm H_2O,mix}=X_{\rm H_2O,eq}$ throughout.  
This behavior indicates that super-Earth-mass planets cannot sustain highly water-enriched envelopes within the disk lifetime, whereas Earth-mass planets can preserve a wider diversity of compositions.  
Hence, the envelope composition at the time of disk dispersal can serve as a strong diagnostic of the planet mass at that epoch.

\medskip
\noindent
\textbf{Estimation of oxygen exhaustion limit} \\
Once the oxygen exhaustion limit is reached, the total amount of water produced is limited by the available reactive oxygen, and the water mass in the vapor-mixed envelope scales approximately as $M_{\rm H_2O,mix} \propto f_{\rm O,react} M_{\rm iso}$, assuming that the fraction of water partitioned into the envelope is nearly mass-independent (as confirmed in our simulations).  
Meanwhile, the hydrogen mass in the same layer is found to scale as $M_{\rm H_2,mix} \propto M_{\rm iso}^{2.5} T_{\rm disk}^{-1.5}$, consistent with previous envelope accretion models~\citep[e.g.,][]{Lee+Chiang2015}.  
Combining these dependences, and calibrating the numerical coefficient with the simulation results, we obtain a semi-analytical expression for the oxygen exhaustion limit:
\begin{align}
    X_{\rm H_2O,mix} &=
    \frac{M_{\rm H_2O,mix}}{M_{\rm H_2,mix}+M_{\rm H_2O,mix}} \notag\\
    &\simeq \qty[1 +
    5.0
    \qty(\frac{M_{\rm iso}}{5\,M_\oplus})^{1.5}
    \qty(\frac{f_{\rm O,react}}{0.1})^{-1}
    \qty(\frac{T_{\rm disk}}{500\,{\rm K}})^{-1.5}
    ]^{-1}.
    \label{eq:O_exhaustion_line}
\end{align}
This relation, plotted as a dashed line in the left panel of Fig.~\ref{fig:Miso_Xenv_nominal}, reproduces the numerical results remarkably well.  

This expression clarifies the physical origin of the oxygen exhaustion limit.  
Higher $f_{\rm O,react}$ increases the total water that can be produced, while higher $T_{\rm disk}$ suppresses hydrogen accretion; both effects raise $X_{\rm H_2O,mix}$ by reducing dilution of the enriched envelope.  
In contrast, increasing planet mass lowers $X_{\rm H_2O,mix}$ because accreted hydrogen mass depends more steeply on $M_{\rm iso}$ than the reactive oxygen reservoir does, leading to stronger dilution.  
Thus, the oxygen exhaustion limit reflects the competition between water production (set by the oxygen budget) and dilution by nebular gas accretion (set primarily by planet mass and disk temperature).
The detailed dependence on these parameters will be discussed in the next section.

\medskip
\noindent
\textbf{At the final state} \\
Comparing the right and left panels, $X_{\rm H_2O,mix}$ remains nearly constant—or slightly decreases—for planets more massive than $\sim2M_\oplus$.  
For lower-mass planets, however, efficient atmospheric escape leads to a substantial post-disk increase in $X_{\rm H_2O,mix}$ (see also Fig.~\ref{fig:tX_Miso}).  
In more massive planets, the envelope mass is large enough that photoevaporation negligibly affects the surface pressure.  
Instead, gradual contraction slightly increases the surface pressure, promoting ingassing of water and causing a modest decline in $X_{\rm H_2O,mix}$ during the post-disk phase.  

Even among low-mass planets, this increase in $X_{\rm H_2O,mix}$ is suppressed for cases that had already reached the oxygen exhaustion limit at $t_{\rm disp}$, since their envelopes are too massive for significant escape.  
As a result, even planets with initially low $X_{\rm H_2O,eq}$ end up with high final $X_{\rm H_2O,mix}$, leading to a reduced compositional diversity among low-mass planets.  
Overall, the final $X_{\rm H_2O,mix}$ values occupy a relatively narrow range across different planet masses, although their absolute levels remain strongly mass-dependent.  
We note, however, that the magnitude of this post-disk evolution depends sensitively on stellar properties, orbital distance, and the adopted photoevaporation model.

\subsection{Parameter Study}
\begin{figure*}
    \centering
    \includegraphics[width=\linewidth]{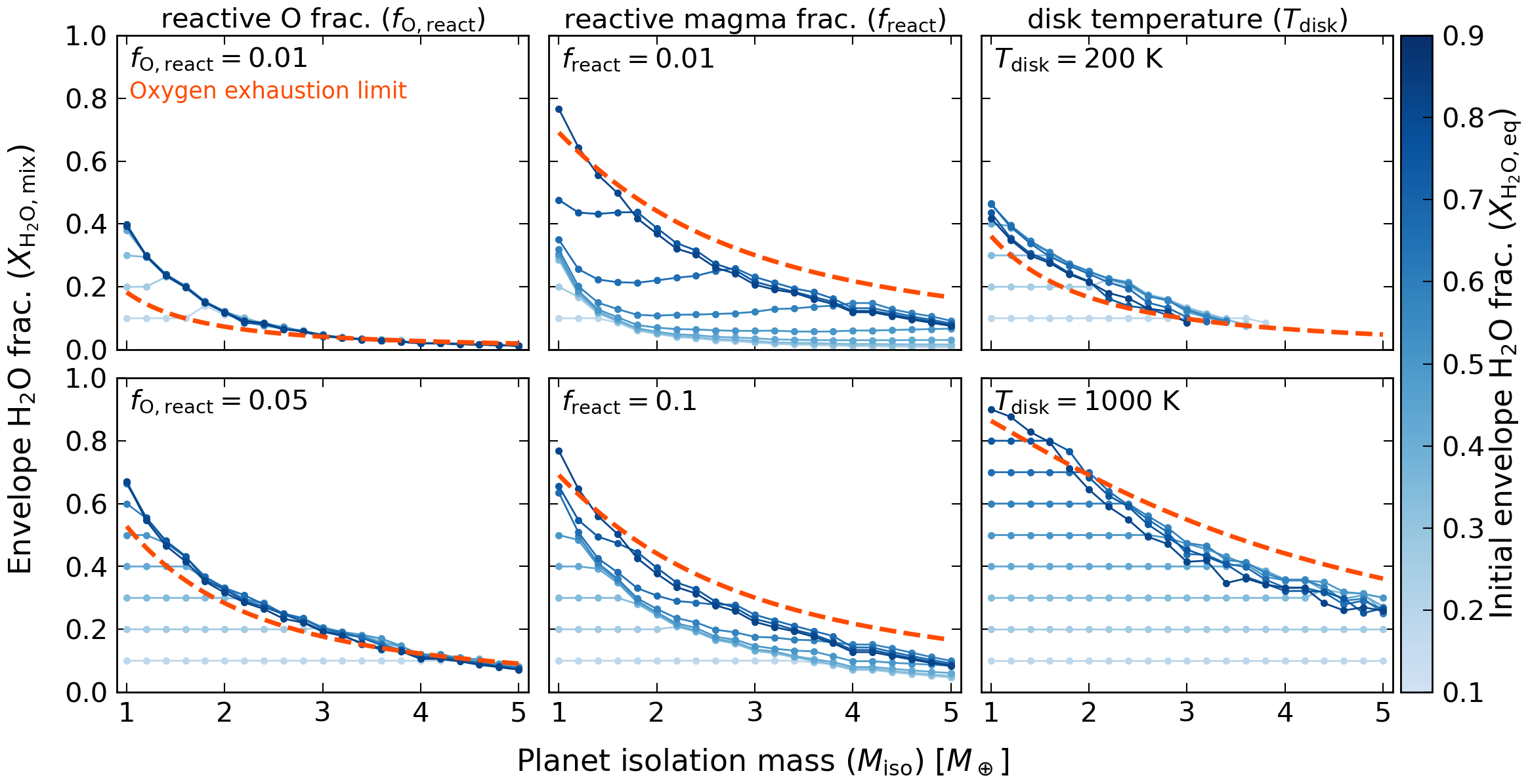}
    \caption{
    Same as Fig.~\ref{fig:Miso_Xenv_nominal}, but for different parameter sets.  
    All panels show the results at the time of disk dispersal.  
    The left, middle, and right columns correspond to variations in $f_{\rm O,react}$, $f_{\rm react}$, and $T_{\rm disk}$, respectively.  
    The dashed lines in each panel represent the semi-analytical oxygen exhaustion limit (Eq.~\ref{eq:O_exhaustion_line}) for each parameter set.  
    Other parameters are fixed to their nominal values.  
    Note that in some runs with $T_{\rm disk}=200$~K, the envelope mass exceeds the core mass, and such planets are excluded from the plots.}
    \label{fig:Miso_Xenv_parameters}
\end{figure*}

Since the resultant envelope composition is controlled by both the efficiency of the water-producing redox reactions and the total amount of accreted disk gas, it is strongly influenced by the magma properties and the surrounding disk conditions.  
To explore these effects, we performed a series of simulations similar to those in Fig.~\ref{fig:Miso_Xenv_nominal}, but varied three key parameters: the mass fraction of reactive oxygen in the magma ($f_{\rm O,react}$), the mass fraction of magma that interacts with the envelope ($f_{\rm react}$), and the nebular temperature ($T_{\rm disk}$).

Overall, the same qualitative trends persist across all parameter sets:  
(1) the upper limit of $X_{\rm H_2O,mix}$---the oxygen exhaustion limit---decreases systematically with increasing planet mass,
and (2) planets that grow to super-Earth masses ($\gtrsim 3M_\oplus$) cannot retain highly water-enriched envelopes at the time of disk dispersal. 
As already described in the previous sections, this robustness arises because the accretion of nebular hydrogen increases quite steeply with planet mass. 
Consequently, once a planet becomes sufficiently massive, hydrogen accretion inevitably dilutes the vapor-mixed layer even when redox reaction is quite efficient.  
However, the location of this oxygen exhaustion limit depends sensitively on the values of $f_{\rm O,react}$, $f_{\rm react}$, and $T_{\rm disk}$, as described below.

\subsubsection{Effects of reactive oxygen abundance ($f_{\rm O,react}$).}
\label{sec:effect_fOreact}
As predicted by Eq.~\eqref{eq:O_exhaustion_line}, decreasing $f_{\rm O,react}$ lowers the oxygen exhaustion limit because the total water yield scales with the available oxygen mass.  
For reference, $f_{\rm O,react}=0.01$ roughly corresponds to an Earth-like FeO content, in which case planets more massive than $\sim2M_\oplus$ end up with nearly hydrogen-dominated envelopes regardless of $X_{\rm H_2O,eq}$.

Compared with Eq.~\eqref{eq:O_exhaustion_line}, our numerical results show slightly higher $X_{\rm H_2O,mix}$ in the low-mass regime ($M_{\rm iso}\lesssim2M_\oplus$) when $f_{\rm O,react}$ is small, while they agree well in the high-mass regime.  
This deviation arises because the total hydrogen mass in the envelope depends on $f_{\rm O,react}$ itself: for larger $f_{\rm O,react}$, $X_{\rm H_2O,mix}$ is higher, promoting more efficient accretion of disk hydrogen. 
Since Eq.~\eqref{eq:O_exhaustion_line} assumes the hydrogen scaling derived from the $f_{\rm O,react}=0.1$ case, it overestimates the hydrogen mass and thus underestimates $X_{\rm H_2O,mix}$ in smaller-$f_{\rm O,react}$ cases.  
In contrast, at higher masses, where the envelope is already hydrogen-dominated even for $f_{\rm O,react}=0.1$, the accreted hydrogen amount hardly depends on $f_{\rm O,react}$, and the semi-analytical line matches the simulations well.

\subsubsection{Effects of the reactive magma fraction ($f_{\rm react}$).}
The influence of $f_{\rm react}$ is more complex, as shown in the middle column of Fig.~\ref{fig:Miso_Xenv_parameters}.  
When $f_{\rm react}<1$, two convergence lines emerge: one followed by the high-$X_{\rm H_2O,eq}$ cases ($\sim0.9$; the upper line) and another by the low-$X_{\rm H_2O,eq}$ cases ($\lesssim0.5$; the lower line).  
Intermediate cases (0.6--0.8) transition from the lower to the upper line as planet mass increases.  
The upper line remains almost unchanged with $f_{\rm react}$ and coincides with the oxygen exhaustion limit for $f_{\rm react}=1$, whereas the lower line shifts downward as $f_{\rm react}$ decreases.

These trends originate from our assumption that $f_{\rm react}=1$ during the solid accretion stage (Phase~I).  
Planets follow the upper line when most of the reactive oxygen is consumed before the end of solid accretion, that is, when vigorous water production occurs early.  
In this case, the total water mass becomes independent of $f_{\rm react}$ after isolation, leading to similar $X_{\rm H_2O,mix}$ values regardless of $f_{\rm react}$.  

Compared to Eq.~\eqref{eq:O_exhaustion_line}, the upper limit is slightly lower in the high-mass region ($M_{\rm iso}\gtrsim2M_\oplus$) because restricted magma–atmosphere contact limits post-exhaustion degassing of water.  
As seen in Fig.~\ref{fig:MRPTWX_nominal}(e), after oxygen exhaustion, the dilution of the vapor-mixed envelope by accreted hydrogen reduces the partial pressure of H$_2$O at the magma surface, promoting degassing from magma.  
However, when $f_{\rm react}<1$, only a fraction of the magma participates in this exchange, so the degassing becomes inefficient and $X_{\rm H_2O,mix}$ decreases slightly.

Conversely, planets follow the lower line when oxygen exhaustion occurs after solid accretion and when water production during the accretion stage is negligible.  
In this regime, the produced water mass scales directly with $f_{\rm react}$.  
If some water is produced during the earlier stage, the resultant $X_{\rm H_2O,mix}$ lies between these two limits.

When most water forms after the termination of solid accretion, the total water mass approximately scales as $\sim f_{\rm O,react}f_{\rm react}M_{\rm iso}$.  
Thus, the cases with $(f_{\rm react},f_{\rm O,react})=(0.1,0.1)$ yield similar total water masses to those with $(1,0.01)$, but $X_{\rm H_2O,mix}$ is higher in the former case because limited reactive magma volume reduces water dissolution, leaving more water in the envelope.

\subsubsection{Effects of disk temperature ($T_{\rm disk}$)}
The nebular temperature regulates gas accretion efficiency by controlling envelope contraction.  
Lower $T_{\rm disk}$ leads to more rapid contraction and thus to more efficient hydrogen accretion, resulting in lower $X_{\rm H_2O,mix}$.  
Conversely, higher $T_{\rm disk}$ suppresses gas accretion, producing a wider diversity of envelope compositions that more directly reflect $X_{\rm H_2O,eq}$.  
Hence, planets formed in warmer inner regions tend to preserve envelope compositions more directly affected by magma chemistry, whereas those in colder outer regions experience stronger dilution by hydrogen-rich gas.  
\rev{Note that even in the cold disk region, increasing the reactive oxygen abundance in magma ($f_{\rm O,react}$) shifts the oxygen exhaustion line upward as discussed in \S~\ref{sec:effect_fOreact}.}
The corresponding exhaustion limits for different $T_{\rm disk}$ are reasonably reproduced by Eq.~\eqref{eq:O_exhaustion_line}.  
Although the hydrogen accretion rate depends non-trivially on $T_{\rm disk}$ via opacity and radiative–convective structure, our semi-analytical formula still provides a good approximation of the upper limit of $X_{\rm H_2O,mix}$ over the explored range of $T_{\rm disk}$ (200–1000~K).
\rev{This sensitivity on the disk temperature also highlights the importance of considering the thermal evolution of disk gas to predict the resultant envelope compositions.}

\subsubsection{Summary}
Both magma properties and disk conditions exert an influence on the resultant envelope mass and composition.  
The upper limit of $X_{\rm H_2O,mix}$, the oxygen exhaustion limit, is mainly governed by the total amount of reactive oxygen available for redox reactions between magma and atmosphere, rather than by the magma redox state alone.  
Across all explored parameter ranges, we consistently find that 
oxygen exhaustion occurs readily for super-Earth-mass planets, producing an exhaustion limit which always trends downward with increasing planet mass.  
This robust trend implies that, regardless of the detailed magma properties or the local disk conditions, retaining a highly water-enriched envelope until disk dispersal is intrinsically difficult once a planet grows to super-Earth mass within the disk lifetime. 

\section{Discussion}

\subsection{Magma Properties and Envelope Composition}

The gas composition produced through reactions between the magma and the atmosphere is strongly dependent on the redox state of the magma~\citep[e.g.,][]{Schaefer+etal2016,Kite+etal2020,Lichtenberg+etal2021,Schlichting+Young2022,Bower+etal2022,Seo+etal2024}.  
Even for magmas buffered near the iron–wüstite equilibrium, the equilibrium mole ratio of H$_2$ to H$_2$O can approach unity, corresponding to a water mass fraction of $X_{\rm H_2O,mix}\sim$0.8--0.9~\citep{Ikoma+Genda2006,Kite+etal2020,Seo+etal2024}.  
Recent experiments support this significant water production under the pressure and temperature condition similar to those at the magma-envelope boundary during formation~\citep{Miozzi+etal2025}.
Other experiments also show that similarly high water fractions can be achieved even in FeO-poor magmas, through the reduction of silicates by metallic iron and hydrogen~\citep{Horn+etal2025}.  
These results predict that such high degrees of water enrichment are thermodynamically favoured under a wide range of conditions.  

If these efficient reactions indeed operate in forming planets, our numerical results imply that oxygen exhaustion would occur almost ubiquitously during the formation stage.  
In that case, the final envelope composition (i.e. the water mass fraction) is primarily controlled by the total amount of reactive oxygen available in the magma and is nearly independent of its initial redox state or convective regime.
This is because the water-producing reaction proceeds so efficiently that the reactive oxygen delivered with the accreting solids is rapidly consumed, often leading to oxygen exhaustion before the end of solid accretion.  
This implies that the present-day envelope composition could serve as a tracer of the compositional properties, and potentially the formation locations, of the planetary core.

However, whether such efficient water production actually occurs depends critically on the dynamics of the magma, particularly the behavior of metallic iron.  
In the case of redox reactions between iron oxides and hydrogen, net oxidation and substantial water formation require the continuous removal of the reduced Fe from the reaction interface.  
Conversely, in more reducing environments, the reduction of silicates under high-pressure conditions demands sufficient metallic Fe as a reactant~\citep{Horn+etal2025}.  
Thus, the persistence and efficiency of the redox reactions at the magma–envelope interface are governed by the transport and segregation behavior of metallic iron within the magma ocean.

The mobility of metallic Fe depends on the turbulence of the magma and on the droplet size distribution.  
Larger planets, with stronger gravity and deeper magma oceans, are expected to exhibit more vigorous turbulence, promoting efficient entrainment and recycling of Fe droplets compared to Earth-sized planets~\citep{Lichtenberg2021,Young+etal2024}.  
On the other hand, the dissolution of water into the magma reduces its density, which can inhibit convection~\citep{Modirrousta-Galian+Korenaga2025}.
The overall efficiency of the redox reactions is therefore determined by a complex interplay between these competing effects.

If the water production is less efficient than suggested by idealised equilibrium models or laboratory experiments, the envelope composition remains below the oxygen-exhaustion limit set by the available reactive oxygen inventory.  
\rev{
Even in this case, our main conclusion remains unchanged: super-Earth-mass planets are unlikely to retain highly enriched envelopes solely through magma--atmosphere interactions.
}
In this regime, however, the envelope composition becomes sensitive to both the precise redox state of the magma and its convective efficiency (Figs.~\ref{fig:Miso_Xenv_nominal} and \ref{fig:Miso_Xenv_parameters}).  
Our simulations show that, in such inefficient cases, the dependence of the final envelope composition on planetary mass becomes weaker than in the oxygen-exhaustion regime.  
This reduced mass dependence may serve as an indicator of whether a planet has experienced vigorous water production through magma–atmosphere interactions.  
Nevertheless, we note that our present model neglects the explicit dependence of the convection state on planet mass, which should be addressed in future studies.

\subsection{Implications for Observations}

\begin{figure}
    \centering
    \includegraphics[width=\linewidth]{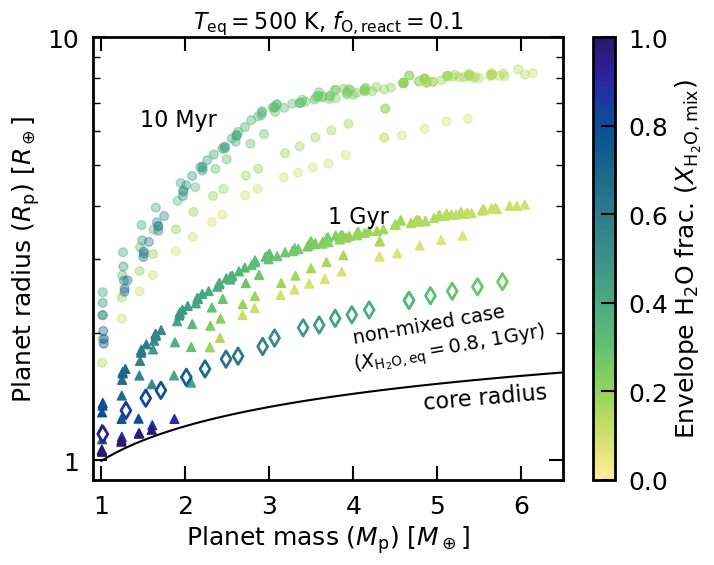}
    \caption{
    Mass--radius relations for simulated planets at 
    10~Myr (faint circles) and 1~Gyr (opaque triangles).
    Each point represents a model with different $M_{\rm iso}$ and $X_{\rm H_2O,eq}$, 
    with other parameters fixed to their nominal values (Table~\ref{tab:parameter}).  
    The colour shows the water mass fraction in the vapor-mixed envelope ($X_{\rm H_2O,mix}$) at each epoch.
    Diamond symbols denote simulations in which the vapor-mixed layer is assumed not to mix with accreted nebular gas after solid accretion ends (shown only for $X_{\rm H_2O,eq}=0.8$; see \S\ref{sec:mixing_assumption}).  
    The black solid line shows the core radius.
    }
    \label{fig:MR_relation}
\end{figure}

\begin{figure}
    \centering
    \includegraphics[width=0.9\linewidth]{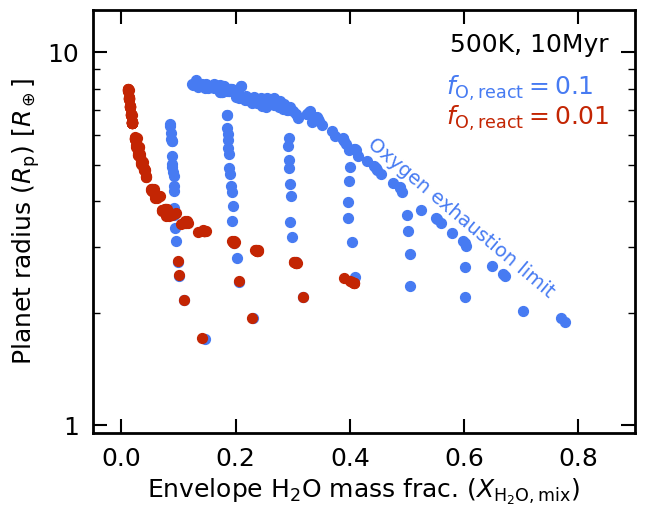}
    \caption{
    Relation between planetary radius ($R_{\rm p}$) and envelope water fraction ($X_{\rm H_2O,mix}$) at 10~Myr. 
    Circles show simulations with various $M_{\rm iso}$ and $X_{\rm H_2O,eq}$.  
    Blue and red points denote cases with $f_{\rm O,react}=0.1$ and 0.01, respectively.  
    Other parameters adopt the nominal values in Table~\ref{tab:parameter}.
    }
    \label{fig:XR_relation}
\end{figure}

Our simulations show that planets growing to super-Earth masses within the disk cannot retain highly water-enriched envelopes: their compositions converge to the oxygen exhaustion limit.  
This outcome is remarkably robust across a broad range of magma and disk parameters.  
Therefore, if the envelope composition of sub-Neptunes is primarily shaped by interactions with magma during formation, their envelopes are unlikely to be extremely metal-rich today.

Nevertheless, planets can acquire highly enriched envelopes through post-disk processes such as giant impacts.  
Formation models suggest that close-in super-Earths frequently undergo multiple giant impacts after disk dispersal~\citep[e.g.,][]{Ogihara+Ida2009,McNeil+Nelson2010,Izidoro+2017}.  
If a planet remains small during the disk phase and grows substantially afterward, it accretes much less nebular hydrogen, making it more likely to retain a water-rich envelope.  
Therefore, the envelope composition of sub-Neptunes may serve as a diagnostic of whether their final mass was assembled before or after the disk phase.

A more direct probe of magma--atmosphere interactions during formation can be provided by young exoplanets.
\rev{
A growing number of planets with ages $\sim$10--100~Myr have recently been detected~\citep{Newton+etal2022,Thao+etal2024,Vach+etal2024,Livingston+etal2026}}, and these systems are less affected by post-disk escape or giant impacts.   
Such systems offer the opportunity to observe the immediate outcomes of envelope formation, including the imprint of the oxygen exhaustion limit.

Figure~\ref{fig:MR_relation} shows the simulated mass--radius relations at 10~Myr and 1~Gyr.  
At each epoch, most planets lie on a single sequence corresponding to cases in which the reactive oxygen in the magma has been exhausted.  
This forms an ``oxygen exhaustion limit'' in the mass--radius plane.  
Planets below this limit correspond to low-$X_{\rm H_2O,eq}$ cases that did not undergo exhaustion.  
Even in the mass range where the envelope becomes nearly hydrogen-dominated, whether oxygen exhaustion occurred still controls the final envelope mass and radius, because vigorous water production promotes more efficient gas accretion during the disk phase, \rev{due to the change in the mean molecular weight and in the adiabat of the envelope gas~\citep{Hori+Ikoma2011,Venturini+2015,Kimura+Ikoma2020}.}
Thus, the magma redox state and the total reactive oxygen content can imprint observable differences in the radii of sub-Neptunes.



When comparing 10~Myr and 1~Gyr, we find that this trend is robust for super-Earths, whereas Earth-mass planets show much larger changes in $X_{\rm H_2O,mix}$ due to post-disk escape and subsequent degassing (see also Fig.~\ref{fig:Miso_Xenv_nominal}).  
As a result, low-mass planets converge to similar compositions by 1~Gyr regardless of their initial $X_{\rm H_2O,eq}$.  
Since the efficiency of post-disk degassing depends on magma convection, observing both young and old low-mass planets in the $\sim$2--4$R_\oplus$ (10~Myr) and $\sim$1--2$R_\oplus$ (1~Gyr) radius ranges would provide constraints on magma–envelope coupling in the post-disk phase.

The young population also shows some observable signatures of the oxygen inventory of the magma.
Figure~\ref{fig:XR_relation} shows the radius–composition relation at 10~Myr.  
As in the mass–radius diagram, an oxygen exhaustion limit appears, and its location varies largely with $f_{\rm O,react}$. 
When $f_{\rm O,react}=0.1$, many young planets lie on this limit and possess inflated radii ($\gtrsim5R_\oplus$) with moderately enriched envelopes ($X_{\rm H_2O,mix}\sim0.1$--$0.4$).  
When $f_{\rm O,react}=0.01$, by contrast, the same inflated planets are almost hydrogen-dominated, and enriched envelopes occur only at much smaller radii.  
Given that \rev{the detection of such inflated young planets with $\sim$10--100~Myrs are recently increasing}, their radii and compositions will provide direct constraints on the oxygen content of their primordial magma oceans, and therefore on the composition and birthplace of the planet.

In summary, our results reveal a clear and quantitative connection between observable properties (mass, radius, and envelope water content) and the geochemical state of the planetary magma during formation.  
Young planets preserve this connection most clearly, while older systems reflect a combination of formation histories and post-disk evolution.  
Future atmospheric characterization of both young and mature sub-Neptunes will therefore provide key constraints on their early volatile budgets, redox states, and formation pathways.

\subsection{Effects of the Mixing Assumption in the Envelope}
\label{sec:mixing_assumption}

The efficiency of magma–envelope interactions and of disk gas accretion depends on how effectively the accreted nebular gas mixes with the underlying enriched layer.  
In our nominal model, we assume efficient mixing within the convective region, identified by the Schwarzschild criterion, neglecting the possible stabilising effects of compositional gradients~\citep{Leconte+etal2017,Selsis+etal2023,nicholls_convective_2024}. 
Although vigorous convection is expected during the high-luminosity accretion phase, a strong compositional gradient may partially suppress convective mixing. 
On the other hand, experimental and theoretical studies show that hydrogen and water are fully miscible under the high-pressure and high-temperature conditions expected in these deep layers~\citep[e.g.,][]{Seward+Franck1981,Bali+etal2013,Vlasov+etal2023,Soubiran+Militzer2015,Bergermann+etal2021,Bergermann+etal2024,Gupta+etal2024}.  
While the long-term evolution of envelopes with composition gradients has been investigated in the contexts of diluted cores of giant planets~\citep[e.g.,][]{Vazan+etal2018a} and post-disk evolution of sub-Neptunes~\citep{Misener+Schlichting2022,Vazan+Ormel2023}, its role during the formation stage remains unexplored.

To examine the effect of the mixing assumption, we performed additional simulations in which the accreted nebular gas does not mix with the vapor-mixed envelope after the termination of solid accretion.  
The diamond symbols in Fig.~\ref{fig:MR_relation} show the resulting mass–radius relation for these non-mixing cases.  
Here we fixed $X_{\rm H_2O,eq}=0.8$ and all other parameters to their nominal values (Table~\ref{tab:parameter}), varying only $M_{\rm iso}$.  
In these simulations, a relatively thick, nebular-composition layer forms atop the vapor-mixed layer, but it quickly escapes after disk dispersal due to the efficient photoevaporation driven by the inflated planetary radius.  
As a result, only the vapor-mixed layer remains at the end of the simulation, and its mass is much smaller than in the efficient-mixing case.  
Consequently, the resulting planetary radii are smaller and the envelope water mass fractions $X_{\rm H_2O,mix}$ tend to be higher in the non-mixing scenario.

We therefore conclude that the degree of mixing between the accreted nebular gas and the enriched envelope has a significant impact on both the envelope mass and the planetary radius--quantities that may be distinguishable by future observations.

\subsection{Caveats and Future Works}

In this study, we have assumed that the magma-atmosphere interaction affects only the water abundance in the envelope.  
However, under the extreme pressure and temperature conditions at the base of the envelope, hydrogen and silicate melts are expected to be fully miscible~\citep[e.g.,][]{Markham+etal2022,Young+etal2024,Stixrude+Gilmore2025a}, which could strongly alter the composition and structure of the deep envelope.  
In addition, reactions between hydrogen and vaporised silicates can generate Si-bearing species such as silane (SiH$_4$)~\citep{Misener+Schlichting2022,Ito+etal2025,Hakim2026MNRAS}.  
As with water, the incorporation of such heavy species substantially modifies the envelope structure and the efficiency of disk gas accretion.  
Because these elements have higher molecular weights and larger reservoirs, their influence could exceed that of water, although the magnitude of this effect depends sensitively on the efficiency of vertical mixing within the envelope.

While we have focused on endogenous enrichment processes, exogenous pathways, such as the accretion of icy planetesimals during the formation phase~\citep{Venturini+2016,Ormel+2021} or direct accretion of disk gas polluted by the sublimation of solids~\citep[e.g.,][]{Booth+2017,Booth+Ilee2019,Schneider+Bitsch2021}, may also contribute.  
\citet{Venturini+2016} found that the co-accretion of disk gas and icy solids can yield envelope water mass fractions up to $\lesssim 50$~wt\% for super-Earth-mass planets, which is higher than the maximum values obtained in our simulations.  
However, their models neglect water dissolution into the magma.  
Thus, a next step will be to investigate how different enrichment mechanisms, endogenous and exogenous, jointly shape the final envelope properties, while self-consistently accounting for water partitioning between magma and atmosphere. 
Such studies will help reveal the formation history of super-Earths and sub-Neptunes from their observed properties.

\section{Conclusions}

We have investigated how magma–envelope interactions during planet formation shape the accumulation, composition, and long-term evolution of primordial atmospheres.  
Our time-dependent model couples disk gas accretion, water production driven by redox reaction, water partitioning between magma and atmosphere, and post-disk thermal evolution and escape.  
This framework allows us to assess how magma properties (redox state, reactive oxygen inventory, and convective efficiency) together with planetary mass control the final atmospheric mass and composition.

Our main findings are as follows:
\begin{itemize}
    \item Water enrichment greatly enhances disk gas accretion, allowing super-Earths to acquire massive envelopes ($\sim$10~wt.\% of the planetary mass) during the disk phase.
    \item When redox reactions proceed efficiently, water production becomes limited by the reactive oxygen inventory in the magma.  
    Once this oxygen exhaustion occurs, continued nebular gas accretion dilutes the envelope, yielding water fractions far below the equilibrium values.
    \item This behaviour produces a clear upper bound on the envelope water mass fraction as a function of planet mass---the {\it oxygen exhaustion limit}.  
    Because gas accretion efficiency increases steeply with planet mass, this limit decreases for more massive planets, making it difficult for planets that reach super-Earth masses during the disk phase to retain highly enriched envelopes by disk dispersal.
    \item The location of the oxygen exhaustion limit depends primarily on the abundance of reactive oxygen in the magma (which sets the maximum water inventory) and on the disk temperature (which regulates hydrogen accretion).  
    In contrast, it is largely insensitive to the magma redox state or convection efficiency, because vigorous water production rapidly consumes the available oxygen regardless of these details.
    \item Observation of young sub-Neptunes ($\lesssim 100$~Myr) can probe the formation-stage magma properties.
    In particular, their radii and inferred envelope compositions can constrain the reactive oxygen inventory of the magma and, by extension, the composition of their building blocks and formation locations.
\end{itemize}

Chemical equilibrium calculations and high-pressure experiments indicate that highly efficient water production---leading to comparable H$_2$ and H$_2$O abundances---is likely common during formation.  
If such conditions are realised, oxygen exhaustion should frequently occur, causing super-Earth and sub-Neptune envelopes to converge toward the exhaustion limit.  
In this regime, the envelope composition becomes a robust indicator of both the magma composition and the planetary mass at disk dispersal, providing strong constraints on formation pathways.

Because both the magma properties and the amount of accreted hydrogen—and thus the oxygen exhaustion limit—are determined by the local disk environment and growth history, integrating these enrichment and partitioning processes into global planet formation models will enable quantitative links between early-stage geochemical and dynamical conditions and present-day observables such as planetary radius and atmospheric composition.

\begin{acknowledgments}
T.K. was supported by JSPS International Leading Research Project (JSPS KAKENHI Grant No. JP22K21344) and Daiichi-Sankyo “Habataku” Support Program for the Next Generation of Researchers.
T.L. was supported by the Branco Weiss Foundation, the Netherlands eScience Center (PROTEUS project, NLESC.OEC.2023.017), the Alfred P. Sloan Foundation (AEThER project, G202114194), NASA’s Nexus for Exoplanet System Science research coordination network (Alien Earths project, 80NSSC21K0593), and the NWO NWA-ORC PRELIFE Consortium (PRELIFE project, NWA.1630.23.013).
Part of the numerical computations were carried out on the Hábrók high performance computing cluster at the University of Groningen. 
\end{acknowledgments}

\begin{contribution}
%

T.K. developed the numerical model and carried out the simulations, based on the idea and advice by T.L.  
Both authors discussed the results and implications and wrote the paper.



\end{contribution}

\bibliography{refer}
\bibliographystyle{aasjournalv7}



\end{document}